\documentstyle[preprint,pra,aps]{revtex}
\begin{document}
\draft
\title{\bf Calculations of parity nonconserving $s$-$d$ transitions 
in Cs, Fr, Ba$^+$, and Ra$^+$}
\author{V.A. Dzuba, V.V. Flambaum, and J.S.M. Ginges}
\address{School of Physics, University of New South Wales, 
Sydney 2052,Australia}
\date{\today}
\maketitle

\tightenlines

\begin{abstract}

We have performed {\it ab initio} mixed-states and sum-over-states 
calculations of parity nonconserving (PNC) electric dipole (E1) 
transition amplitudes between $s$-$d$ electron states of Cs, Fr, 
Ba$^{+}$, and Ra$^{+}$. 
For the lower states of these atoms we have also calculated 
energies, E1 transition amplitudes, and lifetimes.
We have shown that PNC E1 transition amplitudes between $s$-$d$ 
states can be calculated to high accuracy.
Contrary to the Cs $6s$-$7s$ transition, in these transitions there are no 
strong cancelations between different terms in the sum-over-states approach. 
In fact, there is one dominating term which deviates from the sum by less 
than $20\%$. This term corresponds to an $s$-$p_{1/2}$ weak matrix 
element, which can be calculated to better than $1\%$, 
and a $p_{1/2}$-$d_{3/2}$ E1 transition amplitude, which can be measured.
Also, the $s$-$d$ amplitudes are about four times larger than the 
corresponding $s$-$s$ transitions.
We have shown that by using a hybrid mixed-states/sum-over-states 
approach the accuracy of the calculations of PNC $s$-$d$ amplitudes 
could compete with that of Cs $6s$-$7s$ if $p_{1/2}$-$d_{3/2}$ 
E1 amplitudes are measured to high accuracy.

\end{abstract} 
\vspace{1cm}

\section{Introduction}

Precise low-energy experiments on parity nonconservation (PNC) in atoms 
provide a test of the standard model of elementary particle interactions. 
By measuring PNC electric dipole (E1) transition amplitudes, the value 
of the nuclear weak charge can be extracted by comparison with 
calculations. In a recent PNC experiment with cesium \cite{wood} the 
PNC E1 transition amplitude between the $6s$ and $7s$ states has been 
determined with an unprecedented accuracy of 0.35\%. However, 
interpretation of the experiment is limited by 
the accuracy of the atomic calculations. Since 1989, calculations 
of the $6s$-$7s$ transition in Cs have been at the 1\% level 
\cite{flam89pnc,johns90}. At this level of accuracy the value of the 
nuclear weak charge is consistent with that predicted by the Standard Model. 
Recent measurements of values relevant to the PNC E1 amplitude 
(E1 transition amplitudes, hyperfine structure constants) 
are in much better agreement with the calculated values than they were 
ten years ago. From this it has been claimed that the accuracy of the 
calculated PNC E1 amplitude is 0.4\% \cite{bw}. Re-interpreting the Cs
measurement with the higher accuracy, while using the calculations 
\cite{flam89pnc,johns90}, the value of the nuclear weak charge gives a 
$2.5~\sigma$ deviation from the Standard Model prediction \cite{bw}. 
However, inclusion of the Breit interaction into the calculations 
reduces the deviation by about $1~\sigma$ \cite{Der,DzJohns}.  
Note that these measurements give the best limits on new physics beyond 
the Standard Model, such as extra Z bosons, leptoquarks, composite fermions 
\cite{casal,rosner,erler}.

One obviously needs an independent confirmation of the Cs result. 
In this paper we show that the accuracy of calculations of PNC E1 
transitions between $s$-$d$ states of 
Cs, Fr, Ba$^{+}$, and Ra$^{+}$ could compete 
with that of the Cs $6s$-$7s$ transition. 
The experiment for the $6s$-$5d$ transition in Ba$^+$ is 
already in progress \cite{fortson}.

\section{Many-body calculations}

We perform calculations for $N$-electron atoms with one external electron 
above closed shells.
The calculations start from the relativistic Hartree-Fock (RHF) method 
in the $\hat{V}^{N-1}$ approximation. The single-electron RHF Hamiltonian is 
\begin{equation}
\label{eq:RHF}
\hat{H}_{0}=c{\bf \alpha}\cdot \hat{{\bf p}}+(\beta -1)c^{2}-
Z/r+\hat{V}^{N-1},
\end{equation} 
${\bf \alpha}$ and $\beta$ are Dirac matrices and $\hat{{\bf p}}$ is the 
electron momentum.
The accuracy of RHF energies is of the order of $10\%$ for heavy atoms 
like Cs, Fr, Ba$^+$, and Ra$^+$. 

\begin{figure}[b]
\centerline{\psfig{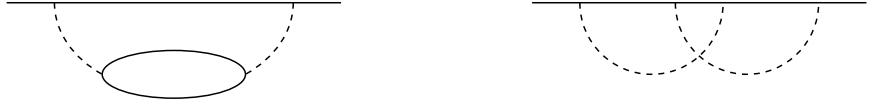}}
\caption{Second-order correlation diagrams for the valence 
electron ($\hat \Sigma$ operator). Dashed line is the Coulomb interaction. 
Loop is the polarization of the atomic core.}
\label{fig:2ndorder}
\end{figure}

In order to obtain more realistic wavefunctions, 
electron-electron correlations must be taken into account.
Correlation corrections to the electron orbitals are calculated 
using the ``correlation potential'' method 
\cite{flam87}. 
This method corresponds to adding a non-local correlation potential 
$\hat{\Sigma}$ to the potential $\hat{V}^{N-1}$ in the RHF equation
(\ref{eq:RHF}) and then solving for the states of the external electron. 
The correlation potential is defined such that its 
average value coincides with the correlation correction to energy, 
$\delta E_{i}=\langle\psi _{i}|\hat{\Sigma}|\psi _{i}\rangle$.
The correlation potential is calculated by means of many-body 
perturbation theory in the residual Coulomb interaction
\begin{equation}
U=\hat{H}-\sum_{i=1}^{N}\hat{H}_{0}({\bf r}_{i})=
\sum _{i<j}^{N}\frac{1}{|{\bf r}_{i}-{\bf r}_{j}|}-
\sum _{i=1}^{N}\hat{V}^{N-1}({\bf r}_{i}),
\end{equation}
where $\hat{H}$ is the exact Hamiltonian of an atom.
The lowest-order correlation diagrams (second-order in $U$) 
are presented in Fig.~\ref{fig:2ndorder}. At this level of calculation 
the accuracy for energy levels is about 1\%.

\begin{figure}[b]
\centerline{\psfig{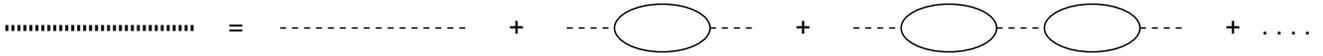}}
\caption{Screening of the Coulomb interaction.}
\label{fig:screening}
\end{figure}

\begin{figure}[b]
\centerline{\psfig{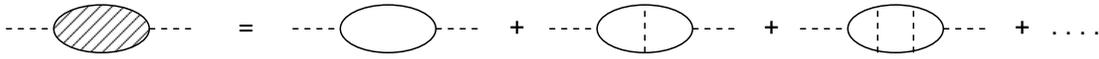}}
\caption{Hole-particle interaction in the polarization operator.}
\label{fig:hpchain}
\end{figure}

\begin{figure}[b]
\centerline{\psfig{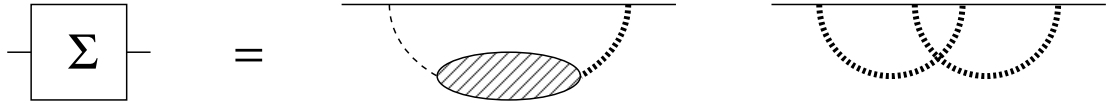}}
\caption{The electron self-energy operator with screening and hole-particle 
interaction included.}
\label{fig:hpscreense}
\end{figure}

\begin{figure}[b]
\centerline{\psfig{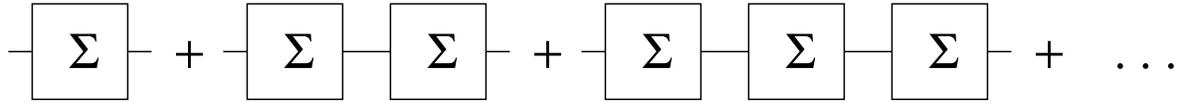}}
\caption{Chaining of the self-energy operator.}
\label{fig:sechain}
\end{figure}

Using the correlation potential method and the Feynman diagram technique 
we include three series of higher order diagrams which are calculated 
in all orders of perturbation theory 
\cite{flam89pnc,flam89energy,flam89e1hfs}. 
These are screening of the electron-electron interaction, the hole-particle 
interaction, and chaining of the self-energy operator $\hat{\Sigma}$. 
The electron-electron screening (see Fig.~\ref{fig:screening}) and the 
hole-particle interaction (Fig.~\ref{fig:hpchain}) are incorporated into 
the self-energy operator $\hat{\Sigma}$ (Fig.~\ref{fig:hpscreense}). 
Chaining of the self-energy operator to all orders 
(Fig.~\ref{fig:sechain}) is then calculated by adding 
$\hat{\Sigma}$ to the Hartree-Fock potential $\hat{V}^{N-1}$ 
and solving the equation
\begin{equation}
(\hat{H}_{0}+\hat{\Sigma}-\epsilon)\psi =0
\end{equation}
iteratively for the states of the external electron.
In this way ``Brueckner'' energies and orbitals are obtained. 
These energies have an accuracy of the order of $0.1\%$.
The wavefunctions can be further modified by placing a coefficient before 
$\hat{\Sigma}$ such that the corresponding energy coincides with the 
experimental value. This fitting of the Brueckner orbitals can be considered 
as a way of including higher-order diagrams into the calculations. 

\begin{figure}[b]
\centerline{\psfig{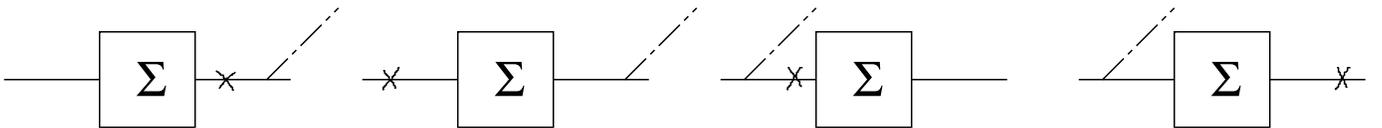}}
\caption{Brueckner-type correlation corrections to the PNC E1 transition 
amplitude; 
the crosses denote the weak interaction and the 
dashed lines denote the electromagnetic interaction.}
\label{fig:pncdom}
\end{figure}

\begin{figure}[b]
\centerline{\psfig{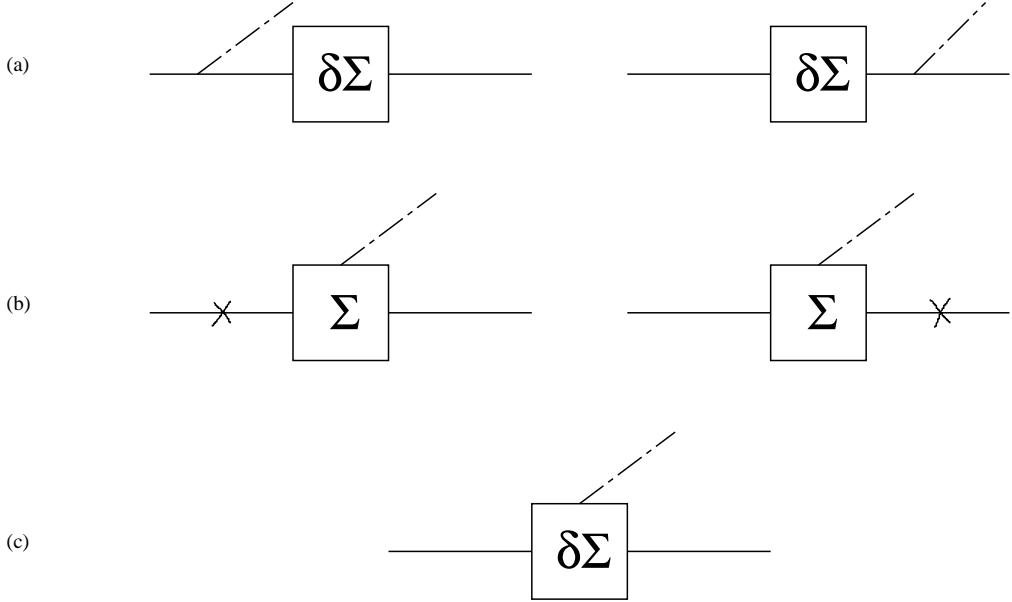}}
\caption{Small corrections to the PNC E1 transition amplitude: 
external field inside the correlation potential.
In diagrams (a) the weak interaction is inside the 
correlation potential ($\delta \Sigma$ denotes the change in $\Sigma $ due to 
the weak interaction); this is known as the weak correlation potential.
Diagrams (b,c) represent structural radiation 
(photon field inside the correlation potential).
In diagram (b) the weak interaction occurs in the external lines;
in diagram (c) both the weak and electromagnetic interactions 
occur in the internal lines.}
\label{fig:pncint}
\end{figure}

We use the time-dependent Hartree-Fock method (which is equivalent to the 
random-phase approximation with exchange) to calculate the interaction of 
external fields with atomic electrons.
In this paper we deal with two external fields: 
the electric field of the photon (E1 transition amplitudes) 
and the weak field of the nucleus.     
In the RHF approximation the interaction between an external field 
$\hat{H}_{\rm ext}$ and atomic electrons is
$\langle \psi ^{HF}_{1}|\hat{H}_{\rm ext}|\psi ^{HF}_{2}\rangle$, 
where $\psi ^{HF}_{1}$ and $\psi ^{HF}_{1}$ are RHF orbitals.  
Inclusion of the polarization of the atomic core by an external field is  
reduced to the addition of a correction $\delta \hat{V}$ 
(which is the correction to the Hartree-Fock potential due to the interaction 
between the core and the external field) 
to the operator which describes the interaction, 
$\langle \psi ^{HF}_{1}|\hat{H}_{\rm ext}+
\delta \hat{V}|\psi ^{HF}_{2}\rangle$.
To include ``Brueckner-type'' correlation corrections the RHF orbitals are 
simply replaced by Brueckner ones, 
$\langle \psi ^{Br}_{1}|\hat{H}_{\rm ext}+
\delta \hat{V}|\psi ^{Br}_{2}\rangle$. 
The Brueckner-type correlations give the dominant corrections to the 
RHF approximation. They correspond to diagrams in which the 
interactions occur in the external lines of the self-energy operator 
(see, e.g., Fig. \ref{fig:pncdom}). 
Those diagrams in which the E1 interaction occurs in the internal 
lines are known as ``structural radiation'', 
while those in which the weak interaction occurs in the internal lines are 
known as the ``weak correlation potential''
(see, e.g., Fig. \ref{fig:pncint}).  
There is also a correction to the amplitudes arising from the 
normalization of states \cite{flam87}. 
The structural radiation, weak correlation potential, and 
normalization contributions are suppressed by the small parameter 
$E_{\rm ext}/E_{\rm core}\sim 1/10$, where  $E_{\rm ext}$ and 
$E_{\rm int}$ are excitation energies of the external and core electrons,
respectively.

The nuclear spin-independent weak interaction of an electron with 
the nucleus is
\begin{equation}
\hat{H}_{W}=-\frac{G_{F}}{2\sqrt{2}}\rho (r)Q_{W}\gamma _{5}
\end{equation}
where $G_{F}$ is the Fermi constant, $Q_{W}$ is the weak charge of the 
nucleus, $\gamma _{5}$ is a Dirac matrix, and $\rho (r)$ is the density 
of the nucleus.
Parity nonconserving E1 transition amplitudes, arising due to the 
simultaneous interaction of atomic electrons with the nuclear weak charge 
and the photon field, can be calculated using two methods:
from a mixed-states approach; 
or from a sum-over-states approach, 
in which experimental values (energies and E1 transition amplitudes) 
can be explicitly included.
Contributions to PNC E1 transition amplitudes are presented diagrammatically  
in Figs. \ref{fig:pncdom}, \ref{fig:pncint}.

\begin{figure}[b]
\centerline{\psfig{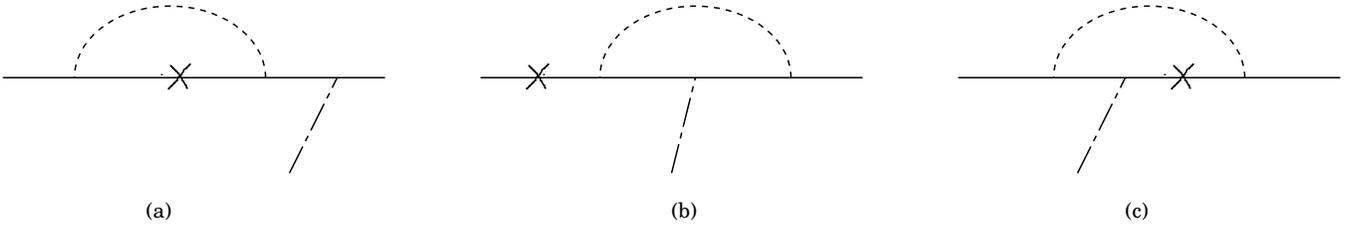}}
\caption{Examples of diagrams representing the polarization of the 
atomic core by external fields. 
(The diagrams we have presented are exchange diagrams; 
there are also direct diagrams.)
In diagrams (a) and (b) the core is 
polarized by a single field (the dashed line denotes the E1 
interaction and the cross denotes the weak interaction). 
Diagram (c) corresponds to the polarization of the core by both fields. }
\label{fig:corepol}
\end{figure}

In the mixed-states approach the PNC E1 transition amplitude between the 
states $ns$ and $(n-1)d_{3/2}$, $n=6$ for Cs and Ba$^{+}$, 
$n=7$ for Fr and Ra$^{+}$, is given by 
\begin{equation}
\label{eq:mixed}
E1_{PNC}=\langle \psi _{(n-1)d}|\hat{H}_{E1}+\delta \hat{V}_{E1}|
\delta \psi _{ns}\rangle +
\langle \psi _{(n-1)d}|\hat{H}_{W}+\delta \hat{V}_{W}|
\tilde{\psi} _{ns}\rangle +
\langle \psi _{(n-1)d}|\delta \hat{V}_{E1W}|\psi _{ns}\rangle ,
\end{equation}
where $\delta \psi $ and $\delta \hat{V}_{W}$ are corrections to 
single-electron wavefunctions and the Hartree-Fock potential caused 
by the weak interaction, $\tilde{\psi}$ and $\delta \hat{V}_{E1}$ are 
corrections to wavefunctions and the Hartree-Fock potential caused 
by the electric field of the photon, $\delta \hat{V}_{E1W}$ is the 
correction to the core potential due to the simultaneous action of the 
weak field and the electric field of the photon;  
the wavefunctions $\psi _{(n-1)d}$ and $\psi _{ns}$ correspond to Brueckner 
orbitals, and the corrections $\delta \psi$ and $\tilde{\psi}$ are found 
by solving the equations:
\begin{eqnarray}
(\hat{H}_{0}+\hat{\Sigma}-\epsilon)\delta \psi &=&
-(\hat{H}_{W}+\delta \hat{V}_{W})\psi \nonumber \\
(\hat{H}_{0}+\hat{\Sigma}-\epsilon)\tilde{\psi}&=&
-(\hat{H}_{E1}+\delta \hat{V}_{E1})\psi  \nonumber .
\end{eqnarray} 
This method is equivalent to calculating the diagrams presented 
in Fig. \ref{fig:pncdom} (with $\hat{\Sigma}$ chained to all orders, 
Fig. \ref{fig:sechain}) with the inclusion of the core polarization 
diagrams presented in Fig.~\ref{fig:corepol}.

Parity nonconserving E1 transition amplitudes between the states
$ns$ and $(n-1)d$ in the sum-over-states approach have the form 
\begin{eqnarray}
\label{eq:sos}
E1_{PNC}&=
&\sum _{n'}\frac{\langle (n-1)d_{3/2}|\hat{H}_{E1}+\delta \hat{V}_{E1}
|n'p_{1/2}\rangle 
\langle n'p_{1/2}|\hat{H}_{W}+\delta \hat{V}_{W}|ns\rangle}
{E_{ns}-E_{n'p_{1/2}}} \nonumber \\ 
&&+
\sum _{n'}\frac{\langle (n-1)d_{3/2}|\hat{H}_{W}+\delta \hat{V}_{W}
|n'p_{3/2}\rangle 
\langle n'p_{3/2}|\hat{H}_{E1}+\delta \hat{V}_{E1}|ns\rangle}
{E_{nd_{3/2}}-E_{n'p_{3/2}}},
\end{eqnarray}
where the sum is taken over a complete set of $p_{1/2}$ and $p_{3/2}$ states.

Note that the sum-over-states approach should also include the states with 
double-excitations like, for example in Cs, 
$\langle 5p^{6}6s|\hat{H}_{W}|5p^{5}6p7s\rangle$.
In the mixed-states approach these states are included, for example, 
in the last term of Eq. \ref{eq:mixed} 
(see also diagram (c) of Fig. \ref{fig:corepol}).
These exotic states contribute due to their mixing 
with the single-excited electron states.
This means that the mixed-states calculation (\ref{eq:mixed}) 
is more complete than the sum-over-states (\ref{eq:sos}) 
{\it unless} the high-energy states with two or more excited electrons 
are included into the sum. 

However, the accuracy of pure {\it ab initio} calculations for $s$-$d$ 
transitions is not very good because of the huge correlations for $d$-states. 
On the other hand, we will see in the next section that this problem can 
be avoided in the sum-over-states approach by using experimental values 
for the $p$-$d$ E1 transition amplitudes. 
Therefore, the best accuracy can be achieved when both methods are combined. 
Substitution of experimental values into the sum-over-states 
approach leads to a correction to the PNC amplitude which can be added to 
the mixed-states result. Following this procedure, it is possible to 
determine the PNC $s$-$d$ amplitudes with an accuracy of about $1\%$ 
(see discussion in the end of the next section). 

\section{Results}

Hartree-Fock energies for Cs, Fr, Ba$^{+}$, and Ra$^{+}$ are presented in 
Table~\ref{tab:energies}. These have an accuracy of the order of 10\%.   
The Brueckner energies, including the three series of higher order diagrams, 
are also presented in Table~\ref{tab:energies}.
These energies have an accuracy of the order of 0.1\%. 

Electric dipole transition amplitudes between the states $m$ and $m'$ are 
calculated in length form, 
$\langle m'||\hat{H}_{E1}||m\rangle = \langle m'||{\bf r}||m\rangle 
=C_{mm'}R$, where $C_{mm'}$ are angular coefficients and 
$R$ is the radial integral. 
In Table \ref{tab:E1} we present radial integrals relevant to the 
sum-over-states calculation for Cs, Fr, Ba$^{+}$, and Ra$^{+}$.
In this table we present the values obtained in the RHF approximation and 
show the contribution of core polarization 
to the RHF integrals; we also present the (unfitted) Brueckner 
results and the contributions arising from structural radiation and 
normalization of states.
In Table \ref{tab:moreE1} we present radial integrals 
between the lower states of the four atoms which are calculated with 
fitted Brueckner orbitals and with structural radiation and normalization 
contributions included. 
Experimental values for Cs and Fr are presented in Table \ref{tab:expe1}.
The Cs transitions $5d_{3/2}$-$6p_{3/2}$ and $5d_{3/2}$-$6p_{1/2}$
were extracted from the measurement \cite{cstau} of the $5d_{3/2}$ lifetime, 
$\tau =909(15)$~ns by assuming that the ratio of the calculated radial 
integrals corresponds to the ratio of the experimental values. 
This assumption was also used to obtain the Fr $7d_{3/2}$-$7p_{3/2}$ and 
$7d_{3/2}$-$7p_{1/2}$ radial integrals from the 
measured lifetime $\tau =73.6(3)$~ns \cite{frdtau}. 
With the exception of the Cs $6s$-$7p_{1/2}$ transition, 
the calculations of $s$-$p_{1/2}$ radial integrals agree with 
experiment at the level of 0.1\%. 
The poor accuracy of the $6s$-$7p_{1/2}$ radial integral is due to the 
fact that the main RHF contribution is very small and the relative 
contribution of all corrections is large. 
The $5d$-$6p$ radial integrals for Cs have poor accuracy, 
deviating from experiment by about 4\%. This is indicative 
of the poor calculation of $d$-states due to very large correlation 
corrections. The accuracy for the Fr $7d$-$7p$ radial integrals is 
about 1\%. The reason that this accuracy is better than that for
Cs $5d$-$6p$ is because the accuracy of the higher $d$-levels 
(here $7d$ rather than $6d$) is better due to smaller correlation 
corrections.

We have calculated the lifetimes of the low-lying states of 
Ba$^{+}$ and Ra$^{+}$.
The $nd_{3/2}$ states of Ba$^{+}$ ($n=6$) and Ra$^{+}$ ($n=7$) decay 
directly to the ground state via the E2 transition; 
the $nd_{5/2}$ states decay via both the E2 and M1 transitions.
Lifetimes of the Ba$^{+}$ $5d$ states are presented in 
Table~\ref{tab:lifetime}.
The calculations were performed with fitted Brueckner orbitals; 
core polarization, structural radiation, and normalization contributions 
were included into the E2 transition amplitudes. 
The calculations are in good agreement with experiment.
We have also presented the calculations performed by 
Guet and Johnson \cite{G&J}.
It appears that for the state $5d_{5/2}$ they have not taken 
into account the M1 transition. This seems to be the reason for the 
discrepancy between the lifetime calculations for this state. 
From our calculations it is seen that inclusion of the M1 transition 
effectively decreases the lifetime of the $5d_{5/2}$ state from $36.3$~s 
to $30.3$~s.      
For Ra$^{+}$ we obtained the lifetimes 
$\tau =0.641$~s and $\tau =0.302$~s for the states
$6d_{3/2}$ and $6d_{5/2}$, respectively; these were calculated in the same 
way as for the Ba$^{+}$ $5d$ states.
The lifetimes for states of Ra$^{+}$ have not been measured.
Lifetimes of all other states are strongly dominated by E1 transitions 
and so can be calculated using the radial integrals presented in 
Table~\ref{tab:moreE1}.
The calculated lifetimes of the $6p$ states of Ba$^{+}$ are in excellent 
agreement with experiment and calculations by Guet and Johnson 
(see Table~\ref{tab:lifetime}).

The mixed-states results for the E1 PNC transition amplitudes are listed 
in Table~\ref{tab:mixed}.
The results of the sum-over-states calculation, 
and the contributions of the six terms corresponding to the summation 
of the $n$-$(n+2)~p$ states, are presented in Table~\ref{tab:sos}.
In both calculations the contributions of structural radiation, 
weak correlation potential, and normalization of states are not included.
The $s$-$d$ PNC amplitudes are up to about four times as 
large as their corresponding $s$-$s$ amplitudes.
Furthermore, unlike the contributions to the sum-over-states calculation 
in Cs $6s$-$7s$, in which the dominant contribution is about twice as large 
as the final result due to strong cancelations between three major terms in 
the sum, the PNC $s$-$d$ transitions in Cs, Fr, Ba$^{+}$, and Ra$^{+}$ 
are strongly dominated by a single term.
In each case this term corresponds to
$\langle (n-1)d_{3/2}|\hat{H}_{E1}|np_{1/2}\rangle 
\langle np_{1/2}|\hat{H}_{W}|ns\rangle /(E_{ns}-E_{np_{1/2}})$;
this term is different from the sum by less than 20\%.

Because the Cs $6p_{1/2}$-$5d_{3/2}$ E1 transition and energies are known
we can correct the mixed-states PNC result. Replacing the calculated 
values by these experimental values in the dominating term of the 
sum (\ref{eq:sos}) for Cs increases this term (and the total sum) by about
$4\%$. This correction is mostly due to the difference between the calculated 
and experimental E1 amplitude (see Table \ref{tab:expe1}).
From the $1\%$ accuracy of calculations of hyperfine structure constants 
for $s$ and $p_{1/2}$ states \cite{flam89e1hfs} we can expect that the accuracy 
of the $s$-$p_{1/2}$ weak matrix elements in this calculation is also about $1\%$.
Therefore, we can say that the uncertainty in the calculated {\it ab initio} 
$s$-$d$ $E1_{PNC}$ amplitudes is dominated by the uncertainty of the $p$-$d$ 
E1 matrix elements and constitutes about $4\%$ for Cs and about the same 
value, or a little more, for other atoms.

The correction to the $6s$-$5d$  $E1_{PNC}$ amplitude in Cs discussed in the 
previous paragraph is $0.126~iea_{B}(-Q_{W}/N)$. When it is added to the 
mixed-states result, the new value is 
$E1_{PNC}(6s-5d)=3.75~iea_{B}(-Q_{W}/N)$.
Since using the experimental $6p$-$5d$ E1 amplitude removes the main source 
of uncertainty, the accuracy of the modified 
result must be considerably better than $4\%$. Assuming high accuracy of 
$p$-$d$ transition amplitudes, one can say that the uncertainty is now 
dominated by the uncertainty of calculated $s$-$p$ weak matrix elements 
which is about $1\%$ (however, we believe that this accuracy can be improved 
beyond $1\%$ with the inclusion of weak correlation potential and 
normalization contributions). 
More rigorous calculations and a more detailed analysis of the accuracy 
will be carried out when the need arises from the progress in experiments. 
We expect that calculations of $E1_{PNC}$ amplitudes for Cs, Fr, Ba$^{+}$, and 
Ra$^{+}$, with empirical corrections, can reach an accuracy of about $1\%$.

\section{Conclusion}

We have calculated the PNC E1 transition amplitudes 
between $s$-$d$ states of Cs, Fr, Ba$^{+}$, and Ra$^{+}$.
Generally, high accuracy cannot be reached in purely {\it ab initio} 
calculations of these transitions due to the poor accuracy of $d$-states.
However, we have shown from a sum-over-states calculation that, unlike 
the Cs $6s$-$7s$ transition, the $s$-$d$ transitions we have mentioned 
are strongly dominated by a single term in the sum.
Moreover, this term corresponds to an $s$-$p_{1/2}$ weak matrix element, 
which can be calculated with an accuracy of better than $1\%$, 
and a $p_{1/2}$-$d_{3/2}$ E1 transition amplitude, which can 
be taken from experiment.
The need to reach high accuracy for $d$ states is therefore avoided.
In addition to this, PNC $s$-$d$ transitions 
are larger than the corresponding $s$-$s$ transitions. 
The mixed-states calculation can be modified by correcting the 
terms in the sum-over-states by inserting experimental E1 transitions and 
energies. If $p_{1/2}$-$d_{3/2}$ E1 transition amplitudes are measured 
to high accuracy, we believe that the accuracy of the calculations of 
$s$-$d$ PNC transitions for Cs, Fr, Ba$^{+}$, and Ra$^{+}$ can reach 1\%.   

\acknowledgments

We are grateful to N. Fortson for useful discussions. 
This work was supported by the Australian Research Council. 


\begin{table}
\caption{Energy levels (ionization potentials) of the lower states of 
Cs, Fr, Ba$^+$ and Ra$^+$ in units $-$cm$^{-1}$.}
\label{tab:energies}
\begin{tabular}{lcccccccc}
 && & Cs & && & Ba$^+$ &\\
State && RHF & Brueckner & Experiment \tablenotemark[1] & 
 & RHF & Brueckner & Experiment \tablenotemark[1]\\
$6s$ && 27954 & 31420 & 31407 && 75339 & 80813 & 80687 \\
$7s$ && 12112 & 12851 & 12871 && 36852 & 38333 & 38332 \\
$8s$ && 6793 & 7082 & 7090 && 22023 & 22651 & 22662 \\ 
$6p_{1/2}$ && 18790 & 20275 & 20228 && 57265 & 60581 & 60425 \\
$7p_{1/2}$ && 9223 & 9643 & 9641 && 30240 & 31332 & 31297 \\
$8p_{1/2}$ && 5513 & 5701 & 5698 && 18848 & 19378 & 19351 \\ 
$6p_{3/2}$ && 18389 & 19708 & 19674& & 55873 & 58860 & 58734 \\
$7p_{3/2}$ && 9079 & 9460 & 9460 && 29699 & 30704 & 30676 \\
$8p_{3/2}$ && 5446 & 5618 & 5615 && 18580 & 19075 & 19051 \\ 
$5d_{3/2}$ && 14138 & 17023 & 16907 && 68139 & 76402 & 75813 \\
$6d_{3/2}$ && 7920 & 8824 & 8818 && 33266 & 34740 & 34737 \\
$7d_{3/2}$ && 4965 & 5362 & 5359 && 20251 & 20871 & 20887 \\ 
$5d_{5/2}$ && 14163 & 16915 & 16810& & 67665 & 75525 & 75012 \\
$6d_{5/2}$ && 7921 & 8781 & 8775 && 33093 & 34536 & 34532 \\
$7d_{5/2}$ && 4963 & 5341 & 5338 && 20167 & 20777 & 20792 \\ 
\hline 
 && & Fr && & & Ra$^+$ & \\
State && RHF & Brueckner & Experiment \tablenotemark[2] & 
 & RHF & Brueckner & Experiment \tablenotemark[1]\\
$7s$ && 28768 & 32841 & 32849 && 75900 & 81960 & 81842 \\
$8s$ && 12282 & 13071 & 13116 && 36861 & 38405 & 38437 \\
$9s$ && 6858 & 7164 & 7178 && 22004 & 22659 & 22677 \\  
$7p_{1/2}$ && 18856 & 20674 & 20612 && 56879 & 60681 & 60491 \\
$8p_{1/2}$ && 9240 & 9730 & 9736 && 30053 & 31244 & 31236 \\
$9p_{1/2}$ && 5521 & 5737 & - && 18748 & 19332 & - \\
$7p_{3/2}$ && 17656 & 18944 & 18925 && 52906 & 55734 & 55633 \\
$8p_{3/2}$ && 8811 & 9180 & 9191 && 28502 & 29447 & 29450 \\
$9p_{3/2}$ && 5319 & 5486 & - && 17975  & 18462 & 18432 \\ 
$6d_{3/2}$ && 13826 & 16610 & - && 62356 & 70149 & 69758 \\
$7d_{3/2}$ && 7725 & 8583 & 8604 && 31575 & 33060 & 33098 \\
$8d_{3/2}$ && 4857 & 5241 & 5248 && 19451 & 20079 & 20107 \\ 
$6d_{5/2}$ && 13925 & 16413 & - && 61592 & 68449 & 68099 \\
$7d_{5/2}$ && 7747 & 8496 & 8516 && 31204 & 32569 & 32602 \\
$8d_{5/2}$ && 4863 & 5197 & 5203 && 19261 & 19849 & 19868 \\  
\end{tabular}
\tablenotetext[1]{Taken from \cite{moore}.}
\tablenotetext[2]{Measured in Refs. 
\cite{Fr1,Fr2,Fr3,Fr4,Fr5,Fr6,Fr7}.}
\end{table}
\begin{table}
\caption{Calculated radial integrals (a.u.) for Cs, Fr, Ba$^+$, Ra$^+$. 
We present RHF values, RHF with core polarization, the Brueckner 
result with core polarization included, and structural radiation 
and normalization of states; 0.0 signifies that the value is smaller 
than the number of figures specified.}
\label{tab:E1}
\begin{tabular}{ccccccc}
Atom &
Transition &
RHF &
RHF + &
Brueckner + &
Structural & 
Normal- \\
 & & & core polar- & core polar- & radiation & ization \\
 & & & ization & ization & & of states\\
\hline
Cs & $6s_{1/2} - 6p_{3/2}$ & -6.432 & -6.074 & -5.500 & 
-0.028 & 0.047 \\
&$6s_{1/2} - 7p_{3/2}$ & -0.602 & -0.440 & -0.463 & 
-0.013 & 0.003 \\
&$6s_{1/2} - 8p_{3/2}$ & -0.245 & -0.143 & -0.162 & 
-0.008 & 0.001 \\
&$6p_{1/2} - 5d_{3/2}$ & 7.775 & 7.481 & 6.050 & 
0.026 & -0.050 \\
&$7p_{1/2} - 5d_{3/2}$ & -3.498 & -3.591 & -1.742 &
0.011 & 0.011 \\
&$8p_{1/2} - 5d_{3/2}$ & -0.860 & -0.916 & -0.556 &
0.007 & 0.003 \\
&&&&&&\\
Fr & $7s_{1/2} - 7p_{3/2}$ & -6.140 & -5.739 & -5.128 &
-0.032 & 0.051 \\  
&$7s_{1/2} - 8p_{3/2}$ & -0.949 & -0.760 & -0.748 &
-0.015 & 0.006 \\ 
&$7s_{1/2} - 9p_{3/2}$ & -0.452 & -0.332 & -0.336 &
-0.009 & 0.003 \\
&$7p_{1/2} - 6d_{3/2}$ & 7.986 & 7.613 & 6.256 & 
0.035 & -0.061 \\
&$8p_{1/2} - 6d_{3/2}$ & -4.005 & -4.116 & -2.249 &
0.014 & 0.015 \\
&$9p_{1/2} - 6d_{3/2}$ & -0.941 & -1.008 & -0.704 &
0.008 & 0.004 \\
&&&&&&\\
Ba$^+$ & $6s_{1/2} - 6p_{3/2}$ & -4.744 & -4.314 & -4.056 &
-0.032 & 0.046 \\ 
&$6s_{1/2} - 7p_{3/2}$ & -0.226 & -0.036 & -0.015 &
-0.013 & 0.0 \\
&$6s_{1/2} - 8p_{3/2}$ & -0.068 & 0.054 & 0.076 &
-0.007 & -0.001 \\
&$6p_{1/2} - 5d_{3/2}$ & 3.244 & 2.964 & 2.634 & 
0.026 & -0.034 \\
&$7p_{1/2} - 5d_{3/2}$ & 0.304 & 0.184 & 0.225 &
0.010 & -0.002 \\
&$8p_{1/2} - 5d_{3/2}$ & 0.169 & 0.092 & 0.099 &
0.007 & -0.001 \\
&&&&&&\\
Ra$^+$ & $7s_{1/2} - 7p_{3/2}$ & -4.624 & -4.154 & -3.885 &
-0.035 & 0.048 \\ 
&$7s_{1/2} - 8p_{3/2}$ & -0.541 & -0.320 & -0.286 &
-0.015 & 0.003 \\
&$7s_{1/2} - 9p_{3/2}$ & -0.243 & -0.098 & -0.067 &
-0.009 & 0.001 \\
&$7p_{1/2} - 6d_{3/2}$ & 3.851 & 3.448 & 3.067 &
0.037 & -0.047 \\
&$8p_{1/2} - 6d_{3/2}$ & 0.091 & -0.075 & 0.009 &
0.014 & 0.0 \\
&$9p_{1/2} - 6d_{3/2}$ & 0.075 & -0.030 & -0.014 &
0.008 & 0.0 \\

\end{tabular}
\end{table}
\begin{table}
\caption{Radial integrals (a.u.) for states of Cs, Fr, Ba$^{+}$ and Ra$^{+}$. 
Fitted Brueckner orbitals are used; core polarization, and 
structural radiation and normalization of states are also 
included.}
\label{tab:moreE1}
\begin{tabular}{cccccccc}
& & $6p_{1/2}$ & $7p_{1/2}$ & $8p_{1/2}$ & $6p_{3/2}$ & 
$7p_{3/2}$ & $8p_{3/2}$ \\
\hline
Cs & $6s$ & -5.508 & -0.313 & -0.081 & -5.482 & -0.471 & -0.171 \\ 
& $7s$ & 5.211 & -12.605 & -1.137 & 5.625 & -12.383 & -1.419 \\
& $8s$ & 1.266 & 11.386 & -21.753 & 1.273 & 12.163 & -21.252 \\
& $5d_{3/2}$ & 6.072 & -1.785 & -0.560 & 
6.120 & -1.579 & -0.506 \\
& $6d_{3/2}$ & -3.696 & 15.558 & -4.234 & -4.082 & 15.612 & -3.726 \\
& $7d_{3/2}$ & -1.806 & -5.632 & 27.670 & -1.918 & -6.379 & 27.773 \\
& $5d_{5/2}$ & - & - & - & 6.190 & -1.641 & -0.522 \\
& $6d_{5/2}$ & - & - & - & -3.986 & 15.699 & -3.886 \\
& $7d_{5/2}$ & - & - & - & -1.893 & -6.186 & 27.876 \\
\hline
Ba$^{+}$ & $6s$ & -4.054 & 0.121 & 0.141 & -4.048 & -0.030 
& 0.063 \\ 
& $7s$ & 3.053 & -8.583 & -0.139 & 3.362 & -8.464 & -0.401 \\
& $8s$ & 0.863 & 6.080 & -14.153 & 0.888 & 6.634 & -13.884 \\
& $5d_{3/2}$ & 2.646 & 0.226 & 0.103 & 2.584 & 0.285 & 0.135 \\
& $6d_{3/2}$ & -4.234 & 7.488 & 0.101 & -4.520 & 7.311 & 0.282 \\
& $7d_{3/2}$ & -1.189 & -7.134 & 13.494 & -1.170 & -7.674 & 13.195 \\
& $5d_{5/2}$ & - & - & - & 2.658 & 0.279 & 0.133 \\
& $6d_{5/2}$ & - & - & - & -4.469 & 7.418 & 0.235 \\
& $7d_{5/2}$ & - & - & - & -1.1865 & -7.5570 & 13.352 \\
\hline
\hline
& & $7p_{1/2}$ & $8p_{1/2}$ & $9p_{1/2}$ & $7p_{3/2}$ & 
$8p_{3/2}$ & $9p_{3/2}$ \\
\hline
Fr & $7s$ & -5.242 & -0.331 & -0.093 & -5.107 & -0.748 & -0.343 \\ 
& $8s$ & 5.217 & -12.326 & -1.206 & 6.484 & -11.536 & -1.947 \\
& $9s$ & 1.256 & 11.428 & -21.382 & 1.215 & 13.777 & -19.660 \\
& $6d_{3/2}$ & 6.237 & -2.229 & -0.691 & 6.417 & -1.553 & -0.516 \\
& $7d_{3/2}$ & -3.014 & 15.893 & -5.488 & -4.186 & 16.175 & -3.829 \\
& $8d_{3/2}$ & -1.606 & -4.296 & 27.952 & -1.975 & -6.546 & 28.464 \\
& $6d_{5/2}$ & - & - & - & 6.576 & -1.684 & -0.549 \\
& $7d_{5/2}$ & - & - & - & -3.974 & 16.368 & -4.184 \\
& $8d_{5/2}$ & - & - & - & -1.917 & -6.106 & 28.695 \\
\hline
Ra$^{+}$ & $7s$ & -3.948 & 0.108 & 0.142 & -3.877 & -0.294 
& -0.082 \\
& $8s$ & 3.104 & -8.523 & -0.168 & 4.038 & -8.071 & -0.891 \\ 
& $9s$ & 0.867 & 6.167 & -14.122 & 0.897 & 7.825 & -13.137 \\
& $6d_{3/2}$ & 3.074 & 0.011 & -0.011 & 2.913 & 0.245 & 0.110 \\
& $7d_{3/2}$ & -3.774 & 8.262 & -0.412 & -4.662 & 7.801 & 0.256 \\
& $8d_{3/2}$ & -1.240 & -6.152 & 14.649 & -1.242 & -7.812 & 13.854 \\
& $6d_{5/2}$ & - & - & - & 3.109 & 0.224 & 0.105 \\
& $7d_{5/2}$ & - & - & - & -4.515 & 8.087 & 0.117 \\
& $8d_{5/2}$ & - & - & - & -1.281 & -7.476 & 14.276 \\
\end{tabular}
\end{table}
\begin{table}
\caption{Calculated (Table \ref{tab:moreE1}) and experimental 
radial integrals (a.u.).}
\label{tab:expe1}
\begin{tabular}{cccc}
Atom & Transition & Calc. & Exp.\\
\hline
Cs & $6s$-$6p_{1/2}$ & -5.508 & -5.497(8) \tablenotemark[1] \\
 & $6s$-$6p_{3/2}$ & -5.482 & -5.476(6) \tablenotemark[1] \\
 & $6s$-$7p_{1/2}$ & -0.313 & -0.348(3) \tablenotemark[2] \\
 & $7s$-$6p_{1/2}$ & 5.211 & 5.185(27) \tablenotemark[3] \\
 & $7s$-$6p_{3/2}$ & 5.625 & 5.611(27) \tablenotemark[3] \\
 & $7s$-$7p_{1/2}$ & -12.605 & -12.625(18) \tablenotemark[4]\\
 & $7s$-$7p_{3/2}$ & -12.383 & -12.401(17) \tablenotemark[4]\\
 & $5d_{3/2}$-$6p_{1/2}$ & 6.072 & 6.31(5)\tablenotemark[5] \\
 & $5d_{3/2}$-$6p_{3/2}$ & 6.120 & 6.36(5)\tablenotemark[5] \\ 
 & $5d_{5/2}$-$6p_{3/2}$ & 6.190 & 6.40(2)\tablenotemark[5] \\
\hline
Fr & $7s$-$7p_{1/2}$ & -5.242 & -5.238(10) \tablenotemark[6]\\ 
 & $7s$-$7p_{3/2}$ & -5.107 & -5.108(13) \tablenotemark[6]\\
 & $7d_{3/2}$-$7p_{1/2}$ & -3.014 & -3.05(1) \tablenotemark[7] \\
 & $7d_{3/2}$-$7p_{3/2}$ & -4.186 & -4.24(2) \tablenotemark[7] \\
 & $7d_{5/2}$-$7p_{3/2}$ & -3.974 & -4.02(8) \tablenotemark[7] \\
\end{tabular}
\tablenotetext[1]{Reference \cite{tanner99}.}
\tablenotetext[2]{Reference \cite{shab79}.}
\tablenotetext[3]{Reference \cite{bouchiat84}.}
\tablenotetext[4]{Reference \cite{brw99}.}
\tablenotetext[5]{Reference \cite{cstau}.}
\tablenotetext[6]{Reference \cite{frptau}.}
\tablenotetext[7]{Reference \cite{frdtau}.}
\end{table}
\begin{table}
\caption{Lifetimes of low-lying states of Ba$^+$.}
\label{tab:lifetime} 
\begin{tabular}{ccccc}
Atom & State & $\tau ^{\rm this ~ work}$ & 
$\tau $\tablenotemark[1] & $\tau ^{\rm expt}$ \\
\hline
Ba$^{+}$ & $6p_{1/2}$ & $7.89$~ns & $7.99$~ns & $7.90(10)$~ns 
\tablenotemark[2] \\
	 & $6p_{3/2}$ & $6.30$~ns & $6.39$~ns & $6.32(10)$~ns 
\tablenotemark[2] \\
	 & $5d_{3/2}$ & $81.5$~s & $83.7$~s & $79.8(4.6)$~s 
\tablenotemark[3]\\
         & $5d_{5/2}$ & $30.3$~s & $37.2$~s & $34.5(3.5)$~s \tablenotemark[4]
\end{tabular}
\tablenotetext[1]{Reference \cite{G&J}.} 
\tablenotetext[2]{Reference \cite{pinn95}.}
\tablenotetext[3]{Reference \cite{N&D}.}
\tablenotetext[4]{Reference \cite{MS90}.}
\end{table}

\begin{table}
\caption{Mixed-states results for PNC E1 transition amplitudes 
between the states $ns$-$(n-1)d$, $\langle (n-1)d|E1_{z}|ns\rangle$, 
where $n=6$ for Cs and Ba$^{+}$ and $n=7$ for Fr and Ra$^{+}$; units are
$10^{-11}iea_{B}(-Q_{W}/N)$.}
\label{tab:mixed}
\begin{tabular}{ccccc}
 & Cs & Fr & Ba$^{+}$ & Ra$^{+}$ \\
\hline
Mixed-states $E1_{PNC}$ & 3.62 & 57.1 & 2.17 & 42.9 \\
\end{tabular}
\end{table}


\begin{table}
\caption{Results of the sum-over-states calculations for the 
PNC E1 transition amplitudes for Cs, Fr, Ba$^{+}$, and Ra$^{+}$.
We present the contributions of the terms in the sum corresponding 
to the intermediate $n$-$(n+2)$$p$ states, the contribution due to 
all other intermediate $p$-states, and the total value; 
units $10^{-11}iea_{B}(-Q_{W}/N)$; 0.0 means that the term is 
smaller than the number of figures specified.}  
\label{tab:sos}
\begin{tabular}{cccc}
& & $\frac{\langle d|\hat{H}_{E1}|np_{1/2}\rangle 
\langle np_{1/2}|\hat{H}_{W}|s\rangle}{E_{s}-E_{np_{1/2}}}$ &
$\frac{\langle d|\hat{H}_{W}|np_{3/2}\rangle 
\langle np_{3/2}|\hat{H}_{E1}|s\rangle}{E_{d}-E_{np_{3/2}}}$ \\
\hline
Cs & n=6 & 3.154 & 0.728 \\
   & n=7 & -0.258 & -0.013 \\
   & n=8 & -0.047 & -0.002 \\
   & Other & \multicolumn{2}{c}{0.197}\\
\hline
   & Total & \multicolumn{2}{c}{3.76}\\
\hline
\hline
Fr & n=7 & 59.78 & 5.19 \\
   & n=8 & -6.13 & -0.15 \\
   & n=9 & -1.10 & -0.03 \\
   & Other & \multicolumn{2}{c}{1.95}\\
\hline
   & Total & \multicolumn{2}{c}{59.5}\\
\hline
\hline
Ba$^{+}$ & n=6 & 2.036 & -0.264 \\
   & n=7 & 0.045 & -0.001 \\
   & n=8 & 0.012 & 0.0 \\
   & Other & \multicolumn{2}{c}{0.511}\\
\hline
   & Total & \multicolumn{2}{c}{2.34}\\
\hline
\hline
Ra$^{+}$ & n=7 & 40.69 & -2.33 \\
   & n=8 & 0.11 & -0.05 \\
   & n=9 & 0.02 & -0.01 \\
   & Other & \multicolumn{2}{c}{7.47}\\
\hline
   & Total & \multicolumn{2}{c}{45.9}\\
\end{tabular}
\end{table}

\center
\widetext
\input psfig
\psfull

\end{document}